# Trust, Experience, and Innovation: Key Factors Shaping American Attitudes About AI

Risa Palm, Justin Kingsland and Toby Bolsen

**Abstract**

A large survey of American adults explored the complex landscape of attitudes towards artificial intelligence (AI). It explored the degree of concern regarding specific potential outcomes of the new advances in AI technology and correlates of these concerns. Key variables associated with the direction and intensity of concern include prior experience using a large language model such as ChatGPT, general trust in science, adherence to the precautionary principle versus support for unrestricted innovation, and demographic factors such as gender. By analyzing these relationships, the paper provides valuable insights into the American public's response to AI that are particularly important in the development of policy to regulate or further encourage its development.

**Introduction**

Artificial intelligence is defined as "a machine-based system that can, for a given set of human-defined objectives, make predictions, recommendations, or decisions that impact real or virtual environments" (National Artificial Intelligence Act of 2020, H.R. 6216). According to this definition, AI systems leverage both machine and human inputs to (a) perceive real and virtual environments, (b) transform these perceptions into models through automated analysis, and (c) use these models to generate options for information or action. Currently, AI is employed in a range of applications including mapping technologies, handwriting recognition for mail sorting, spam filtering, language translation, financial trading, and more.

Artificial intelligence (AI) is rapidly evolving and has the potential for profound impacts. It promises to enhance efficiency and effectiveness across numerous tasks, potentially leading to the creation of new industries and high-paying jobs. In healthcare, AI could facilitate earlier detection of diseases such as cancer and heart disease, accelerate drug discovery, and lower costs while improving access to treatments for previously incurable conditions. Additionally, advanced AI systems in facial recognition and predictive algorithms could enhance community safety by preventing fraud and identifying suspicious activities.

However, there are also significant concerns about the negative impacts of AI. It could result in widespread job losses, as robots and AI systems replace workers across various fields, even those traditionally held by "white collar" workers and managers such as accountants, lawyers, and doctors. In healthcare, the rush to market AI applications without sufficient testing could lead to mistakes, medical errors, and misdiagnoses. Advanced facial recognition technology could threaten community safety and security by influencing decisions about policing and emergency response, potentially worsening social inequalities. Additionally, while AI could transform education, the widespread use of tools like ChatGPT might erode students' writing skills and reduce meaningful interactions with human teachers, leading to diminished face-to-face engagement and interpersonal socialization.

Although prior research has examined Americans' general attitudes toward AI, there has been limited systematic analysis of the characteristics associated with these attitudes and the underlying reasons. This paper aims to explore the nature of Americans' concerns about AI in detail and to connect these concerns with broader attitudes about science and innovation, ultimately providing a deeper understanding necessary to inform public policy about its governance.

**Primary Dimensions of Users Concerns about AI**

Potential applications of AI are nothing short of transformational and are also the subject of "concern."  Popular and academic literature has outlined some of these concerns, ranging from sentient robots that threaten the future of humankind to the possibilities that artificial intelligence could supplement or replace the work of humans in providing education, guiding police, or conducting war.  In a recent best-selling book Mustafa Suleyman (2023) writes: "AI, synthetic biology and other advanced forms of technology produce tail risks on a deeply concerning scale. They could present an existential threat to nation-states—risks so profound they might disrupt or even overturn the current geopolitical order".  Raymond Kurzweil (2024) writes that: "We need to recognize the fact that AI technologies are inherently dual-use… For instance, the very same drone that delivers medication to a hospital that is inaccessible by road during a rainy season could later carry an explosive to that same hospital..."  Like Suleyman, he focuses on the need for "alignment:" how to make the AI systems safe and aligned with humanity's wellbeing. Bengio et al. [4] noted that progress in AI could result in disastrous outcomes: "Increases in capabilities and autonomy may soon massively amplify AI's impact, with risks that include large-scale social harms, malicious uses, and an irreversible loss of human control over autonomous AI systems... AI systems threaten to amplify social injustice, erode social stability, enable large-scale criminal activity, and facilitate automated warfare, customized mass manipulation, and pervasive surveillance."

Surveys indicate that most Americans have heard at least a little about AI but are unfamiliar with many of its specific applications and have limited experience with generative AI technologies (Ballard, 2024; Beets et al., 2023; Dupont et al., 2024; Faverio & Tyson, 2023). Eom et al. (2024) summarized 14 national opinion surveys conducted in the US about AI between 2019 and 2024, supplementing their findings with an additional survey they conducted. The authors concluded that there is increased discourse around the technology, but that "the hype about AI often conflates or confuses the definition of the technology, making it difficult to ascertain whether the public has a singular, shared understanding". National surveys suggest that the U.S. public views AI with a mix of optimism and apprehension, recognizing both its potential benefits and threats (Brewer et al., 2022; Zhang and Dafoe, 2020). This paper adds to this rapidly growing literature, but also provides a more nuanced analysis of how a few factors such as experience with generative AI technology, general attitudes towards trust in science, general attitudes about the need to innovate without boundaries as opposed to hesitate until the side-effects are better understood, and a set of demographic variables associated with specific concerns about the impacts of the technology on individuals and society.

**Primary Correlates of General Attitudes Towards AI**

A variety of correlates of response to individual applications of AI as well as the general notion of AI itself have been explored in previous surveys. In a frequently cited paper, Zhang and Dafoe (2020) found that older Americans are less supportive of developing AI, while those familiar with the technology are more supportive [see also Calice et al., 2022). Beets et al. (2023) found significant differences between Black and White respondents, with Black respondents tending to be more concerned about AI increasing discrimination based on health risks. They also found that women were more concerned than men about using robotic nurses to diagnose and administer medicine to bed-ridden patients. Borwein et al. (2024) found significant differences between men and women in their attitudes towards automation and AI in the workplace. Similarly, Dupont et al. (2024) found age/gender differences in the percentages of those using large language models with males aged 18-44 most likely to be regular users, and males and females over age 55 least likely to have used such models. Many other specific findings of correlates can be cited, but clearly there is a need for more interpretation of likely correlates providing a more nuanced understanding. What our analysis attempts to do is to integrate the various correlates that previous survey research has found to be associated to try to discern theoretically motivated patterns.

### *Prior Experience with AI*

In general, prior experience with a technology has been found to shape intentions and subsequent behavior. As outlined in their general Theory of Planned Behavior, Fishbein and Ajzen (2002) argued that the attitude towards a behavior and therefore the likelihood that it will result in a behavioral change is a function of "accessible beliefs regarding the behavior's likely consequences" (Ajzen, 2020). This set of belief is itself related to the individual's experience with the behavior. Another general model of technology acceptance is the technology acceptance model or TAM (Davis 1989; Davis 1993). This model has been commonly used to assess the acceptance of information technology (Taylor & Todd, 1995) and posits that prior experience is linked to such predictors of technology acceptance as ease of use, accessibility in the memory, and perceived usefulness.

With respect specifically to AI, sophistication about the algorithms on which artificial intelligence is based has been found consistently to influence attitudes about the technology and assessments of its risks or benefits. For example, in a series of studies in Germany, Said et al. (2024) found that a higher level of knowledge about AI was associated with an assessment of both lower risks and more benefits. Horowitz and his colleagues (2024) explored the ways in which experience and familiarity with AI conditions support for various AI-enabled systems such as autonomous vehicles, autonomous weapons systems (AWS), autonomous surgery, as well as personal adoption of AI. They found that except for the support for AWS, prior knowledge and experience was positively associated with a greater willingness to use and support these AI systems. In a set of surveys comparing adult members of the US public with graduate students in a computer science or analytics program, O'Shaughnessy et al. (2023) found that the students, in other words those with more experience with and knowledge about AI, were more "confident and positive in their attitudes towards AI" and more likely to support its general use. In general, then, greater knowledge about AI as indicated by **prior experience** has been linked with increasingly positive general attitudes about AI policy and individual use. Therefore, we pose this hypothesis:

**Hypothesis 1:** *Individuals' who report higher levels of **Prior Experience** using an AI tool (i.e., ChatGPT) will: (a) express lower levels of general concern about AI user applications, (b) express greater support for the future development of AI, (c) express increased beliefs that the benefits of AI outweigh the risks; and, (d) view the positive aspects of AI being greater than the potential negatives.*

***Trust in Science***

A robust literature has linked the idea of trust in science with the likelihood of adoption of new technologies (Bolsen et al., 2014; Spaccatini et al., 2022; Bayes et al., 2023; Dohle et al., 2020; Peng, 2022). Several studies have made the explicit link between trust in science or scientists and attitudes towards AI. Yang et al. (2023) noted the conceptual difference between trust in the technology itself, based primarily on its usefulness and effectiveness, as opposed to trust in scientists or institutions developing the technology, based on "perceptions of the underlying intentions of and value alignment with the institution." Thus, for example, surveys can investigate the "trust" that people have in AI based on a combination of performance (what the technology does) and purpose (why it was developed). Respondents are likely to exhibit more trust when they are confident about the performance of the technology and have "ownership" or freedom of choice in using the technology (see, for example, Kim & Lee, 2024). Yang et al.'s work (2023) advanced this idea by focusing not on trust in the effectiveness of the AI, but instead trust in the scientists and companies that develop and promote it. They examined trust in the scientists as well as technology companies "to act responsibility" in developing AI. They found that support for the development of AI as well as the balance of perceived benefits as opposed to perceived risks was related to trust in the scientists. Along these lines, but distinct from this prior research, our paper sought to measure the relationship between general trust in the effects of science on society: that science enables us to overcome almost any problem or that science creates unintended consequences and replaces older problems with new ones.

**Hypothesis 2:** *Individuals' who report higher levels of **Trust in Science** will: (a) express lower levels of general concern about AI user applications, (b) express greater support for the future development of AI, (c) express increased beliefs that the benefits of AI outweigh the risks; and, (d) view the positive aspects of AI being greater than the potential negatives.*

*Innovation vs. the Precautionary Principle*

As we have already noted, new technologies arouse not only excitement and anticipation, but also fears and desires to control the speed and path of innovation. A major tension surrounding artificial intelligence is centered on its regulation: should innovation at any speed and cost be encouraged or should developers be subject to regulations based on the idea that "if one is embarking on something g new, one should think very carefully about whether it is safe or not, and should not go ahead until reasonably convinced it is" (Saunders, 2000 as quoted in Andorno, 2024). This is the fundamental belief behind the "precautionary principle," first codified into policy at the 1972 Conference on the Human Environment related to the topic of environmental sustainability (Read & O'Riordan, 2017). There is significant controversy at present about how development of future applications artificial intelligence should be regulated, with those at one end of the spectrum advocating a strict application of the "precautionary principle," limiting unfettered experimentation while those at the other end calling for an "innovation principle," promoting development with as little regulation as possible (Hemphill, 2020; Thierer, 2023).

David et al. (2024) examined the beliefs of US adults about the governance of AI given both trust in the stakeholders (individuals, corporations and governments) and trust in the AI itself. They found that those with greater trust in technology corporations preferred greater corporate self-regulation.

To examine the relationship between views on innovation and views about AI, O'Shaughnessy and his colleagues (2023) found that "techno-skepticism", a construct constructed from phrases such as "new technologies are more about making profits rather than making people's lives better" and "I feel uncomfortable about new technologies", was a strong predictor of support for the use of AI applications. Therefore, we anticipate that **general optimism about technology** as well as the **cautions about unregulated development** will be beliefs that are strongly associated with individuals' general concerns and support for the development of AI.

**Hypothesis 3:** *Individuals' who report higher levels of **Optimism toward Innovation (H3a)** will: (a) express lower levels of general concern about AI user applications, (b) express greater support for the future development of AI, (c) express increased beliefs that the benefits of AI outweigh the risks; and, (d) view the positive aspects of AI being greater than the potential negatives.*

Conversely,
**Hypothesis 4:** *Individuals' who report higher levels of **support for the Precautionary Principle** will: (a) express higher levels of general concern about AI user applications, (b) express less support for the future development of AI, (c) express increased beliefs that the risks of AI outweigh the benefits; and, (d) view the negative aspects of AI being greater than the potential positives.*

### Background Characteristics Linked with Support for AI

Survey research has also identified several demographic characteristics often associated with general attitudes about artificial intelligence. The relationships between these characteristics and support for artificial intelligence development are sometimes inconsistent and may be related simply to prior experience with or familiarity with the AI application, or more general societal attitudes. The characteristics most typically identified include gender and age, although other factors are also identified (Kreps et al., 2023).

*Gender*

Numerous surveys have revealed gender differences in attitudes toward artificial intelligence (Ismatullaev & Kim, 2024; Armutat et al., 2024). These differences may stem from factors such as lower participation rates of women in engineering programs and companies that develop AI, along with perceptions of power inequality and discrimination in AI applications (West et al., 2019). Howington (2023) noted that women are adopting AI more slowly than men and tend to be less optimistic about its impact in the workplace. Additionally, West et al. (2019) found that women express greater concern than men about the potential negative effects of AI on their children and personal security. A survey of women in the U.S. conducted by Reading et al. (2023) revealed worries about "unintended medical harm" and "inappropriate data sharing." In a comparative study on attitudes towards workplace automation and AI, Borwein et al. (2024) identified significant differences between men and women regarding perceptions of AI fairness.

*Age*

Survey research has consistently found a relationship between age and attitude towards and use of information technology including applications using artificial intelligence (Hargittai & Dobransky, 2017; Köttl et al., 2021). Ballard (2024) reported that Americans' general attitudes about AI are correlated with age, with younger people more likely to trust AI than their elders. Most Americans over age 45 use neither text generation nor chatbots, and less than 20% stated that they believe that AI is making their lives easier. Wilson et al. (2023) [conducted an in-depth study of the experience of the use of digital devices and social media by older adults in Britain, documenting that barriers to use included both perceived skills (those who perceived themselves to be unskilled in using digital media were reluctant to use it) and physical barriers (problems of eyesight, hearing or dexterity interfering with the use of these technologies).

Stypinska (2023) argues that not only are there differences in uptake by age, but that AI exhibits significant age bias: "(1) age biases incorporated in algorithms and digital datasets (technical level), (2) age stereotypes, prejudices, and ideologies of actors in the field of AI (personal/actor level), (3) invisibility or clichéd representations of category of age and old age in discourses around AI (discourse level), (4) discriminatory effects of use of AI technology on older age groups (group level), (5) exclusion as users of AI technology, services and products (user level)."

*Religiosity*

Several studies have undertaken an analysis of the influence of **religious attitudes** on perception and response to artificial intelligence. In a study of Polish university students, Kozak and Fel (2024) found that people with strong religious beliefs "appear to be more susceptible to feelings of fear regarding AI than their non-religious or religiously neutral counterparts". Using survey data across 68 nations, Jackson et al. (2023) found that although exposure to sciences has little or no effect on personal religiosity, exposure to automation, the use of robots and artificial intelligence, may result in a decline in religion as "automation allows humans to "break" laws of nature, gives humans "superhuman" abilities, and allows humans to "do things that we have never been able to do before." Another survey (Karatas and Cutright, 2023) found that "thinking about God leads people to be more willing to accept recommendations from AI systems than they otherwise would."

*Race*

Race has been an inconsistent predictor of response to AI. In a survey of "AI uptake" where the respondent could choose a human provider or an AI computer system, Black respondents were less likely to choose the AI (Robertson et al, 2023). In contrast, Schiff et al., (2023) found no differences between White and Black respondents with respect to "trust, willingness to share data, and willingness to pay taxes" to support predictive policing.

*Party Identification*

Party *identification* may also influence the general beliefs and opinions of Americans about AI. For example, Yang et al. (2023) found an interaction between political ideology and both trust in scientists and trust in AI: "conservatives were less trusting in scientists" and "political ideology also moderated the effects of trust on support for AI." Similarly, Castelo and Ward (2021) also found that political conservatism was associated with a distrust in AI and aversion to its adoption.

*Education and Income*

Both education and income have sometimes been related to AI attitudes in empirical surveys but have not shown consistent patterns. Given the scattered associations of these background variables, some more consistently related than others, we posed the following

**Research Question**:

*Do gender, age, race, religiosity, education, party identification, and income impact Americans' general concerns and opinions regarding AI after accounting for more proximate beliefs and identities? (i.e., see H1-H4, above).*

**Description of Survey**

To explore the areas of concern about AI and the correlates of these attitudes, we examine the results of a large, nationally representative survey of 1,330 US adults conducted in May 2024. The survey was conducted by Forthright Access a privately managed company, and their panel is proprietary and managed by Bovitz, Inc. Although it is not a probability sample, it can be understood as "nationally representative" in that it is matched to census benchmarks for age, gender, education, Census region and race (Mernyk et al., 2022). [50]. As an overview of our respondents, 68 percent were white, 14 percent were African American, 48 percent were female, the mean age was 47, and the median income was in the $40-49,000 category, similar to national data as reported by the U.S. Census of Population

(https://www.census.gov/quickfacts/fact/table/US/PST045222). All respondents were given a brief definition of artificial intelligence at the beginning of the survey: "Artificial intelligence is the ability of computers to learn and act like humans, such as recognizing patterns, making decisions, and solving problems." They were then asked a series of questions about the specific concerns and opinions discussed above regarding AI.

**Measures:** *Primary Dependent Variables*

We focus on four primary dependent variables in the analyses that follow, including level of concern about specific AI applications, support for the future development of AI, belief that the benefits of AI outweigh the risks or vice versa, and beliefs about the positive and negative impacts of AI on humans and societies. The survey included 8 items/questions that measured various concerns about AI applications, derived from some of the issues commonly raised in popular and academic analysis of AI. Specifically, we asked respondents: "How concerned or worried are you about the possible following outcomes of the further development of artificial intelligence?" (1= Not at all concerned, 2= A little bit concerned, 3= Somewhat concerned, 4= Moderately concerned, 5= Concerned, 6= Very concerned, 7= Extremely concerned) (Table 1). In the analyses below, we also employed a scaled measure (alpha = .87) for *Concerns about AI*.

> **Table 1. Concerns about AI**
>
> a. Your health care provider is an AI robot and recommends treatment for your cancer diagnosis.
> b. You have trained for your profession and find it satisfying. But your job is threatened by implementation of AI
> c. Realistic photos and audio clips are being circulated to try to influence an election with false information.
> d. People are manipulated into giving up person information such as their social security numbers or credit cards by AI generated phone calls or messages.

---

[1] The full set of sample descriptives statistics is provided in the Appendix.

> e. Using facial recognition technology and predictive algorithms to anticipate where crime will occur, police will double-down on certain communities leading to over-policing.
> f. Kindergarten children are taught to read by robots that look like cute furry kittens. The children come to see these robots as their friends and prefer to talk to them than with other children or adults.
> g. In the not-too-distant future, robots could become sentient or conscious.
> h. Robots could become caregivers for the elderly.

We measured three additional general attitudes beyond user concerns about artificial intelligence with the following questions: "To what extent do you oppose or support the further development of artificial intelligence?" (1= strongly oppose; 7=strongly support) (*Support Development*); "Considering the further development of artificial intelligence, do you think the benefits outweigh the risks, or the risks outweigh the benefits? (1=risks definitely outweigh the benefits; 7=benefits definitely outweigh the risks) (*Benefits Outweigh Risks*); ***AI Impact Beliefs (Positive / Negative)*** were measured with five items that asked respondents to report (on a scale of 1–7 where 1 is definitely negative and 7 is definitely positive) the degree to which "future advancements in AI" would have "a more negative or positive impact on: (1) jobs, (2) healthcare, (3) public safety, (4) education, and (5) people's quality of life" (alpha = .92).[2]

*Measures: Primary Independent variables*

We measured **prior experience** with AI by asking respondents: "How frequently have you used an AI writing tool such as ChatGPT?" (1= never; 2= once a week; 3= 2-3 times a week; 4= 4-6 times a week; 5= daily). We assessed participants' **trust in science** with the following question, "Do you think that science enables us to overcome almost any problem or that science creates unintended consequences and replaces older problems with new ones?" (1–7 scale where 1 is definitely overcomes problems and 7 is definitely creates new problems). We measured the **precautionary principle** that scientists should fully understand potential issues before advancing new AI applications by asking the extent to which respondents disagreed or agreed with the following statement: "Scientists should understand all of the problems that might arise in the future before releasing more AI applications." (1–7 scale where 1 is strongly disagree and 7 is strongly agree). We measured belief in the **innovation principle** by asking the extent to which respondents disagreed or agreed with the following statement: "The innovators of Silicon Valley should be given the freedom to innovate and progress while thinking and planning intelligently for any collateral effects of AI" (1= strongly disagree; 7= strongly agree). Religiosity was measured with the following question: "How religious would you say you are?" on a scale of 1 (not at all religious) to 7 (very religious). We also included standard measures for gender, age, race, political party, income and education.

---

[2] To examine the reliability of the *AI Impact Beliefs* scale, we conducted a factor analysis using the 5 constituent measures. The results of the power analysis, including Eigenvalues and factor loadings for each constituent item, are reported in the supplementary appendix. We find a single factor emerges (Eigenvalue=3.805).

**Results**

The overall level of "concern" with each of the specific outcomes of AI was generally high. The average for all eight specific concerns was 4.45 - - close to the midpoint between "moderately concerned" and "concerned" on the 7-point scale. We find that respondents were most concerned with the possibility that realistic photos and audio clips created with false information using artificial intelligence could be used to manipulate an election (Figures 1 and 2). Almost 40 percent of the respondents answered that they were "extremely concerned" with this outcome of AI and another 18 percent said they were "very concerned"; less than 10 percent were "not at all" or only "a little bit" concerned.

The second greatest type of concern was the possibility that people could be manipulated into giving up personal information by AI generated phone calls or messages. Here, slightly more than ten percent showed no or little concern, while just over 50 percent said they were either "very" or "extremely" concerned.

At the other end of the scale, respondents were concerned but less so about healthcare being provided by robots (32% were very or extremely concerned), or robots providing care for the elderly (less than 30% were very or extremely concerned).

**Figure 1. Level of Concern about Specific Aspects of AI**

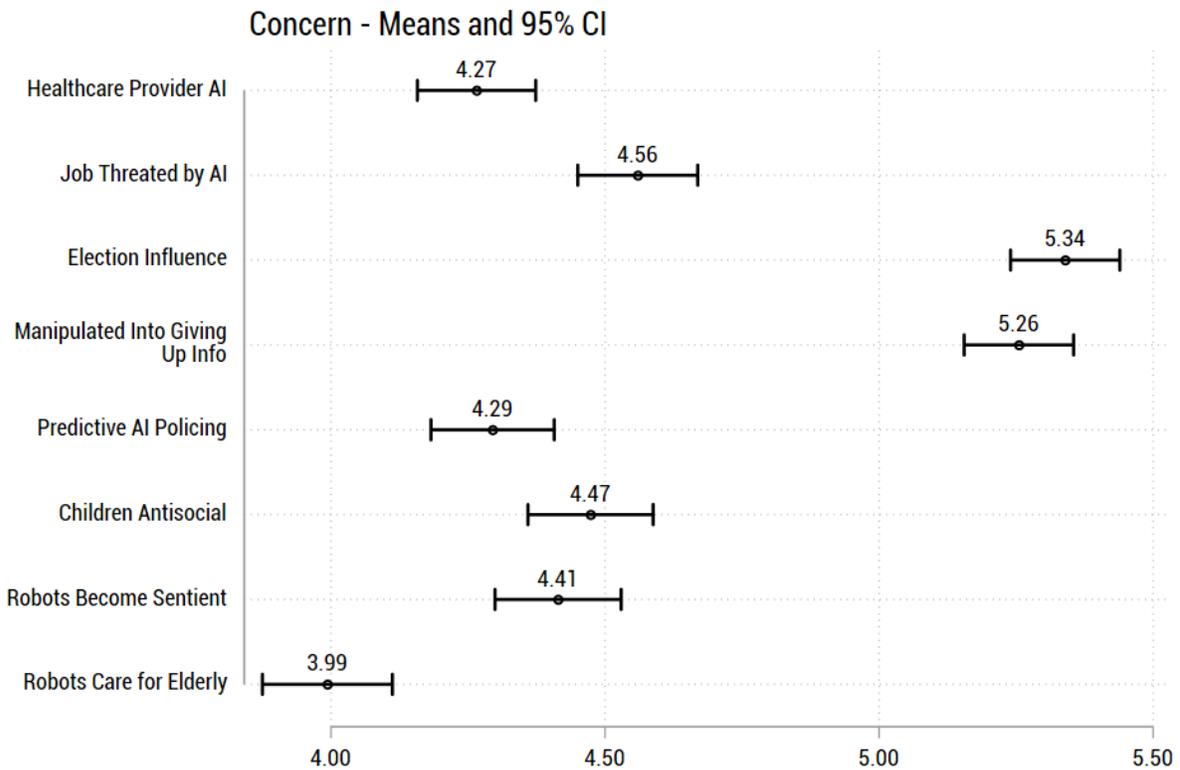

**Figure 2. Distributions of Respondents' Concerns about Specific Aspects of AI**

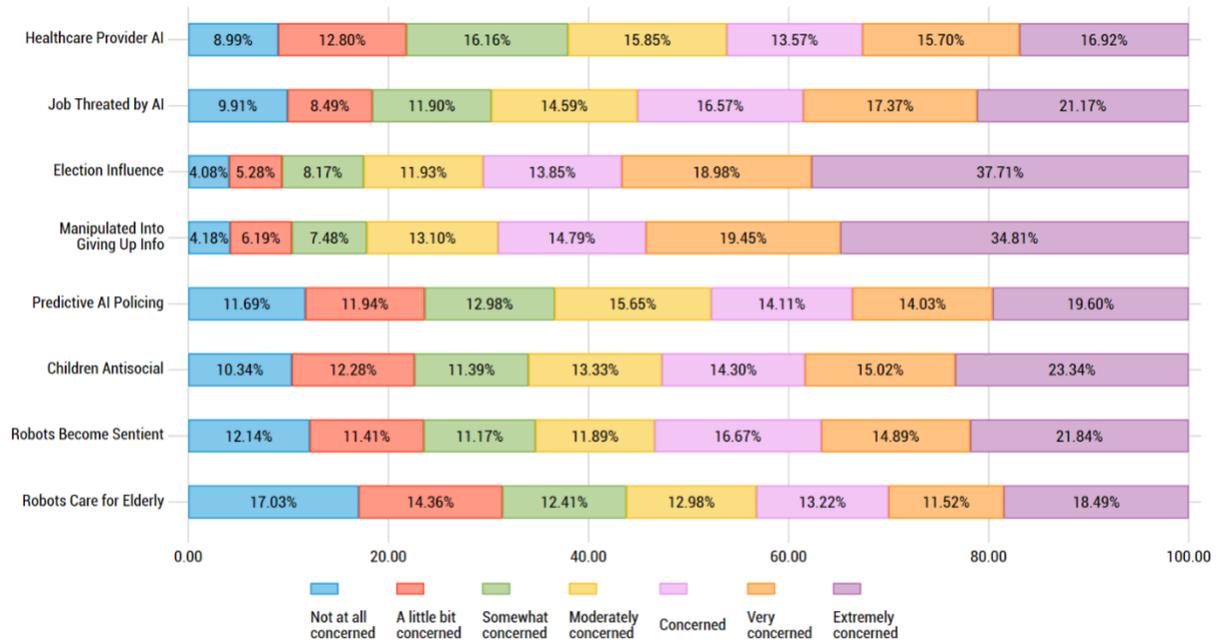

*Multivariate Analysis*

To test our proposed hypotheses, we estimate a series of Ordinary Least Squares models regressing each dependent variable on our predictors of interest. Table 2 reports the model estimates, with cell entries containing OLS coefficient estimates and robust standard errors in parentheses below. To aid the interpretation of the results, we present the point estimates with bars representing a 95% confidence interval in Figure 3 below.

**Table 2:** Determinants of AI Concerns, Support for Its Development, and General Beliefs

|  | (1) Concern Scale Coef. | p-value | (2) Support Development Coef. | p-value | (3) Benefits Outweigh Risks Coef. | p-value | (4) Negative or Positive Effects Scale Coef. | p-value |
|---|---|---|---|---|---|---|---|---|
| Used AI Tool | -0.26*** (0.08) | 0.001 | 0.86*** (0.08) | 0.000 | 0.65*** (0.09) | 0.000 | 0.53*** (0.07) | 0.000 |
| Trust in Science | -0.17*** (0.03) | 0.000 | 0.22*** (0.03) | 0.000 | 0.22*** (0.03) | 0.000 | 0.23*** (0.02) | 0.000 |
| Precautionary Principle | 0.22*** (0.03) | 0.000 | -0.07** (0.03) | 0.020 | -0.16*** (0.03) | 0.000 | -0.10*** (0.02) | 0.000 |
| Silicon Valley Innovate | -0.17*** (0.03) | 0.000 | 0.38*** (0.03) | 0.000 | 0.36*** (0.03) | 0.000 | 0.31*** (0.02) | 0.000 |
| Female | 0.27*** (0.08) | 0.001 | -0.36*** (0.08) | 0.000 | -0.28*** (0.08) | 0.001 | -0.27*** (0.07) | 0.000 |
| White | -0.06 (0.09) | 0.495 | 0.00 (0.09) | 0.977 | -0.01 (0.09) | 0.874 | -0.01 (0.07) | 0.840 |
| Education | -0.03 (0.02) | 0.205 | 0.08*** (0.02) | 0.000 | 0.11*** (0.02) | 0.000 | 0.05*** (0.02) | 0.010 |
| Income | 0.04** (0.02) | 0.014 | -0.03 (0.02) | 0.165 | -0.02 (0.02) | 0.416 | -0.02 (0.02) | 0.231 |
| Republican | 0.03 (0.09) | 0.741 | 0.08 (0.10) | 0.436 | 0.12 (0.10) | 0.238 | 0.01 (0.08) | 0.866 |
| Democrat | -0.01 (0.09) | 0.887 | 0.05 (0.10) | 0.640 | 0.09 (0.10) | 0.349 | 0.15* (0.08) | 0.062 |
| Religiosity | 0.08*** (0.02) | 0.000 | -0.01 (0.02) | 0.522 | 0.01 (0.02) | 0.586 | 0.03* (0.02) | 0.084 |
| Constant | 4.45*** (0.25) | 0.000 | 1.67*** (0.26) | 0.000 | 1.63*** (0.26) | 0.000 | 1.83*** (0.21) | 0.000 |
| N | 1189 |  | 1192 |  | 1188 |  | 1193 |  |
| AIC | 3991.9 |  | 4103.0 |  | 4131.9 |  | 3653.7 |  |
| BIC | 4052.9 |  | 4164.0 |  | 4192.9 |  | 3714.7 |  |

Note: Cell entries are OLS coefficients with standard errors in parentheses below; p-values are shown in the adjacent column. Stars indicate a statistically significant coefficient estimate using a two-tailed test. *p<0.10, **p<0.05, ***p<0.01

Consistent with *Hypothesis 1*, we find that *Prior Experience* using an AI tool such as ChatGPT was associated with lower levels of concern about AI applications (*H1a*; $b=-0.26$, $p<0.001$), greater support for future development (*H1b*; $b=0.86$, $p<0.01$), a belief that the benefits outweigh the risks (*H1c*; $b=0.65$, $p<0.01$), and the beliefs about the impacts of AI being more positive than negative (*H1d*; $b=0.53$, $p<0.01$).

Assessing the influence of *Trust in Science* on AI beliefs, the model estimates provide support for *Hypothesis 2*. Individuals with higher levels of *Trust in Science*, or greater optimism about the impact of scientific and technological development for society, express lower levels of general concern about AI user applications (*H2a*; $b=-0.17$, $p<0.01$) and greater support for the development of AI technologies (*H2b*; $b=0.22$, $p<0.01$). Further, higher levels of reported Trust in Science are associated with a significant increase in beliefs that the benefits of AI outweigh the risks (*H2c*; $b=0.22$, $p<0.01$) and beliefs that AI will have greater positive impacts than negative (*H2d*; $b=0.23$, $p<0.01$).

**Figure 3: Determinants of AI Concerns, Support for Its Development, and General Beliefs**

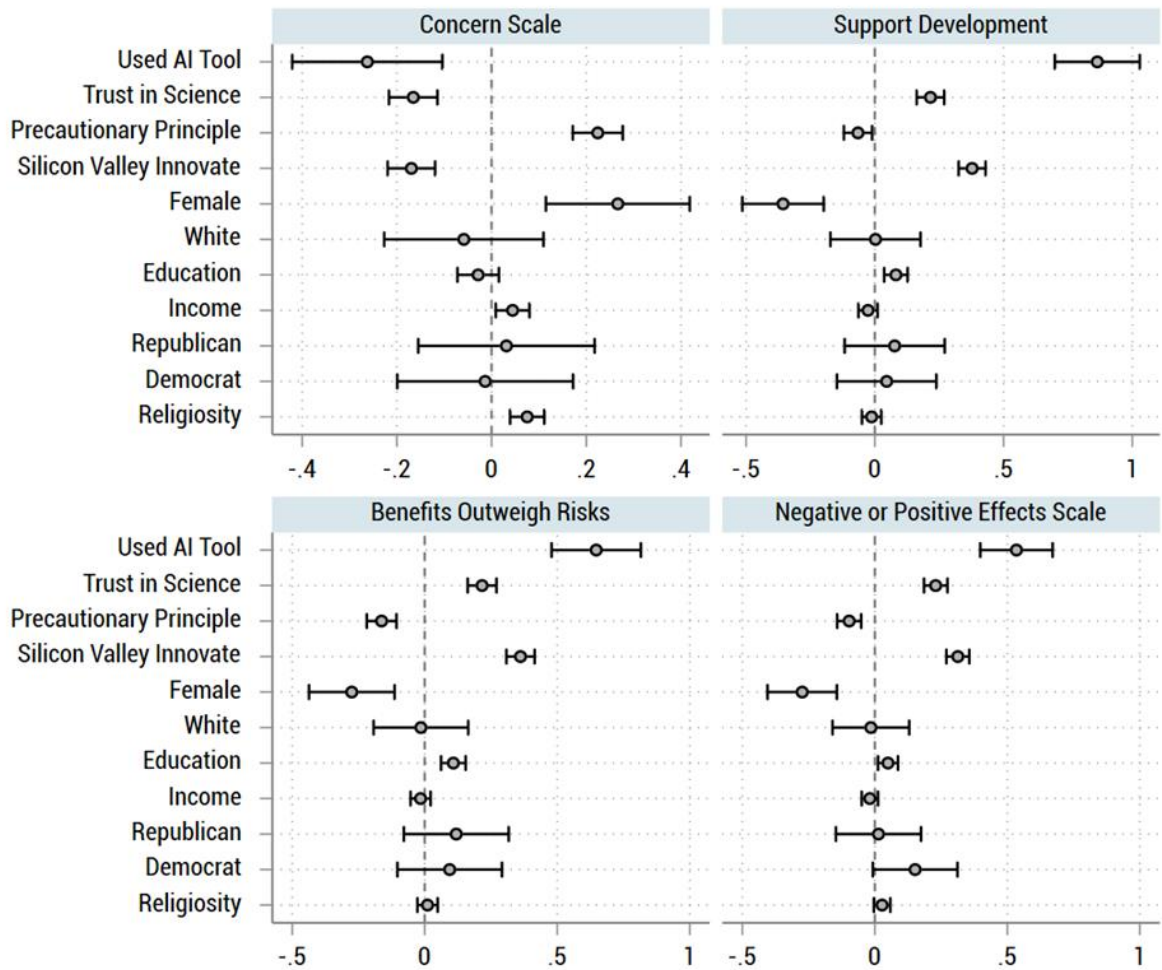

*Note:* Dots are OLS coefficient estimates with error bars representing a 95% confidence interval.

Our final set of hypotheses relate to predictions about the influence of belief in the innovation and precautionary principles on AI beliefs. Specifically, we hypothesized that higher agreement with the innovation principle, that the Silicon Valley scientists should be left to innovate with relatively few regulations, would correspond with lower levels of concern (H3a), and increased support for future development of AI (H3b), belief that benefits of AI outweigh the risks (H3c), and positive impact perceptions (H3d). Conversely, we predicted that higher levels of agreement with the Precautionary Principle would show the opposite profile: higher levels of concern (H4a), reduced support (H4b), beliefs that the risks of AI outweigh the benefits (H4c), and negative impact perceptions (H4d).

The results of the analysis are consistent with our empirical predictions from Hypothesis 3 and Hypothesis 4. Higher levels of agreement with the innovation principle (Silicon Valley Innovate) is associated with lower levels of concern ($b=-0.17$, $p<0.01$), increased support for development ($b=0.38$, $p<0.01$), belief that benefits outweigh risks ($b=0.36$, $p<0.01$), and more positive impact perceptions ($b=0.31$, $p<0.01$) while higher agreement with the Precautionary Principle has an opposite effect: more concern ($b=0.22$, $p<0.01$), less support for development ($b=-0.07$, $p<0.01$), perceive risks outweighing benefits ($b=-0.16$, $p<0.01$), and more negative perceptions about the effects of AI ($b=-0.10$, $p<0.01$).

Other co-variates were more inconsistent in their association with these dependent variables. Gender, that is, self-identification as a female, was related to greater concern, opposition to development, the belief that risks outweigh benefits and a negative outlook on the effects of AI. The other independent variables were related to some but not all the dependent variables. Higher levels of education were linked with supporting development, viewing the benefits as outweighing the risks and a more positive view of the effects of AI. Greater religiosity and higher income were correlated with greater general concern. However, self-identifying as "White", Republican or Democrat was related to none of the dependent variables. In short, only gender was consistent in its relationship to the dimensions of concern, support for development, views of risks versus benefits and views about negative or positive effects, suggesting that among the various correlates, gender merits further attention in future research designs.

**Discussion and Conclusions**

A large-scale survey of a panel representative of the US population, given only a minimal definition of AI, found consistent patterns of concerns with the technology and correlates of these concerns. Respondents expressed significant concern about various applications of AI, particularly regarding its potential to manipulate elections with false information. Additional worries included AI-generated scams and the role of AI in healthcare.

Despite these concerns, many participants supported the continued development of AI, especially those with prior experience using AI tools like ChatGPT. A considerable number believed that the benefits of AI outweighed its risks, particularly in fields such as healthcare and employment.

Factors influencing these attitudes included prior experience with AI, which was associated with lower concern and greater support for its development. Trust in science also played a role; those who had more faith in scientific advancements tended to be less worried about AI and more supportive of its growth. The tension between innovation and precautionary principles was evident, with support for innovation linked to reduced concern and increased optimism about AI. In contrast, adherence to precautionary principles led to heightened concern and skepticism regarding AI's benefits.

Demographic influences revealed that gender was a consistent factor, with female respondents generally expressing more concern about AI and opposing its development more than their male counterparts. Other demographic variables, such as education and religiosity, influenced attitudes as well, but their effects were less consistent across the board.

As artificial intelligence evolves, and particularly such technologies as autonomous vehicles, autonomous weapons and humanoid robotics advance, there is likely to be an even greater need to understand public response and the support for various governance options. Future research should investigate the ways in which communication about these technologies influences the nature of public support.

**References**

Ajzen I (2020) The theory of planned behavior: Frequently asked questions. Human Behav and Emerg Tech 2:314–324. https://doi.org/10.1002/hbe2.195

Ajzen I, Fishbein M (2002) Understanding attitudes and predicting social behavior, Transferred to digital print on demand. Prentice-Hall, Englewood Cliffs, NJ

Andorno R (2004) The Precautionary Principle: A New Legal Standard for a Technological Age. Journal of International Biotechnology Law 1:. https://doi.org/10.1515/jibl.2004.1.1.11

Armutat S, Wattenberg M, Mauritz N (2024) Artificial Intelligence – Gender-Specific Differences in Perception, Understanding, and Training Interest. icgr 7:36–43. https://doi.org/10.34190/icgr.7.1.2163

Ballard J (2024) Americans' top feeling about AI: caution. In: YouGov. https://today.yougov.com/technology/articles/49099-americans-2024-poll-ai-top-feeling-caution. Accessed 24 Sep 2024

Bayes R, Bolsen T, Druckman JN (2023) A Research Agenda for Climate Change Communication and Public Opinion: The Role of Scientific Consensus Messaging and Beyond. Environmental Communication 17:16–34.
https://doi.org/10.1080/17524032.2020.1805343

Beets B, Newman TP, Howell EL, et al (2023) Surveying Public Perceptions of Artificial Intelligence in Health Care in the United States: Systematic Review. J Med Internet Res 25:e40337. https://doi.org/10.2196/40337

Bengio Y, Hinton G, Yao A, et al (2024) Managing extreme AI risks amid rapid progress. Science 384:842–845. https://doi.org/10.1126/science.adn0117

Bolsen, T., Druckman, J. N., & Cook, F. L. (2014). How frames can undermine support for scientific adaptations: Politicization and the status-quo bias. *Public Opinion Quarterly*, *78*(1), 1-26.

Borwein S, Magistro B, Loewen P, et al (2024) The gender gap in attitudes toward workplace technological change. Socio-Economic Review 22:993–1017.
https://doi.org/10.1093/ser/mwae004

Borwein, S., Magistro, B., Loewen, P., Bonikowski, B., & Lee-Whiting, B. (2024). The gender gap in attitudes toward workplace technological change. *Socio-Economic Review*, mwae004.

Brewer PR, Bingaman J, Paintsil A, et al (2022) Media Use, Interpersonal Communication, and Attitudes Toward Artificial Intelligence. Science Communication 44:559–592.
https://doi.org/10.1177/10755470221130307

Calice M, Bao L, Newman T, et al (2022) U.S. Public Attitudes on Artificial Intelligence. https://doi.org/10.17605/OSF.IO/K82D6

Castelo N, Ward AF (2021) Conservatism predicts aversion to consequential Artificial Intelligence. PLoS ONE 16:e0261467. https://doi.org/10.1371/journal.pone.0261467

David, P., Choung, H., & Seberger, J. S. (2024). Who is responsible? US Public perceptions of AI governance through the lenses of trust and ethics. *Public Understanding of Science*, 09636625231224592.

Davis FD (1989) Perceived Usefulness, Perceived Ease of Use, and User Acceptance of Information Technology. MIS Quarterly 13:319. https://doi.org/10.2307/249008

Davis FD (1993) User acceptance of information technology: system characteristics, user perceptions and behavioral impacts. International Journal of Man-Machine Studies 38:475–487. https://doi.org/10.1006/imms.1993.1022

Dohle S, Wingen T, Schreiber M (2020) Acceptance and adoption of protective measures during the COVID-19 pandemic: The role of trust in politics and trust in science. Soc Psychol Bull 15:e4315. https://doi.org/10.32872/spb.4315

Dupont J, Baron D, Price A, et al (2024) What Does The Public Think About AI? (UK). Center for Data Innovation

Eom D, Newman T, Brossard D, Scheufele DA (2024) Societal guardrails for AI? Perspectives on what we know about public opinion on artificial intelligence. Science and Public Policy scae041. https://doi.org/10.1093/scipol/scae041

Faverio M, Tyson A (2023) What the data says about Americans' views of artificial intelligence. In: Pew Research Center. https://www.pewresearch.org/short-reads/2023/11/21/what-the-data-says-about-americans-views-of-artificial-intelligence/. Accessed 24 Sep 2024

Hargittai E, Dobransky K (2017) Old Dogs, New Clicks: Digital Inequality in Skills and Uses among Older Adults. Canadian Journal of Communication 42:195–212. https://doi.org/10.22230/cjc.2017v42n2a3176

Hemphill TA (2020) "The innovation governance dilemma: Alternatives to the precautionary principle." Technology in Society 63:101381. https://doi.org/10.1016/j.techsoc.2020.101381

Horowitz MC, Kahn L, Macdonald J, Schneider J (2024) Adopting AI: how familiarity breeds both trust and contempt. AI & Soc 39:1721–1735. https://doi.org/10.1007/s00146-023-01666-5

Howington J (2023) The AI Gender Gap: Exploring Variances in Workplace Adoption. In: FlexJobs Job. https://www.flexjobs.com/blog/post/the-ai-gender-gap-exploring-variances-in-workplace-adoption/. Accessed 24 Sep 2024

Ismatullaev UVU, Kim S-H (2024) Review of the Factors Affecting Acceptance of AI-Infused Systems. Hum Factors 66:126–144. https://doi.org/10.1177/00187208211064707

Jackson JC, Yam KC, Tang PM, et al (2023) Exposure to automation explains religious declines. Proc Natl Acad Sci USA 120:e2304748120. https://doi.org/10.1073/pnas.2304748120Karataş M, Cutright KM (2023) Thinking about God increases acceptance of artificial intelligence in decision-making. Proc Natl Acad Sci USA 120:e2218961120. https://doi.org/10.1073/pnas.2218961120

Karataş, M., & Cutright, K. M. (2023). Thinking about God increases acceptance of artificial intelligence in decision-making. *Proceedings of the National Academy of Sciences*, *120*(33), e2218961120.

Kim S-W, Lee Y (2024) Investigation into the Influence of Socio-Cultural Factors on Attitudes toward Artificial Intelligence. Educ Inf Technol 29:9907–9935. https://doi.org/10.1007/s10639-023-12172-y

Köttl H, Gallistl V, Rohner R, Ayalon L (2021) "But at the age of 85? Forget it!": Internalized ageism, a barrier to technology use. Journal of Aging Studies 59:100971. https://doi.org/10.1016/j.jaging.2021.100971

Kozak J, Fel S (2024) The Relationship between Religiosity Level and Emotional Responses to Artificial Intelligence in University Students. Religions 15:331. https://doi.org/10.3390/rel15030331

Kreps S, George J, Lushenko P, Rao A (2023) Exploring the artificial intelligence "Trust paradox": Evidence from a survey experiment in the United States. PLoS ONE 18:e0288109. https://doi.org/10.1371/journal.pone.0288109

Kurzweil, R (2024) The Singularity is Nearer: when we merge with AI. Viking.

Mernyk JS, Pink SL, Druckman JN, Willer R (2022) Correcting inaccurate metaperceptions reduces Americans' support for partisan violence. Proc Natl Acad Sci USA 119:e2116851119. https://doi.org/10.1073/pnas.2116851119

National Artificial Intelligence Act of 2020, H.R. 6216

O'Shaughnessy MR, Schiff DS, Varshney LR, et al (2023) What governs attitudes toward artificial intelligence adoption and governance? Science and Public Policy 50:161–176. https://doi.org/10.1093/scipol/scac056

Peng, Y. (2023). The role of ideological dimensions in shaping acceptance of facial recognition technology and reactions to algorithm bias. *Public Understanding of Science*, *32*(2), 190-207.

Read R, O'Riordan T (2017) The Precautionary Principle Under Fire. Environment: Science and Policy for Sustainable Development 59:4–15. https://doi.org/10.1080/00139157.2017.1350005

Reading Turchioe M, Harkins S, Desai P, et al (2023) Women's perspectives on the use of artificial intelligence (AI)-based technologies in mental healthcare. JAMIA Open 6:ooad048. https://doi.org/10.1093/jamiaopen/ooad048

Robertson, C., Woods, A., Bergstrand, K., Findley, J., Balser, C., & Slepian, M. J. (2023). Diverse patients' attitudes towards Artificial Intelligence (AI) in diagnosis. *PLOS Digital Health*, *2*(5), e0000237.

Said, N., Wendt, S., & Potinteu, A. E. Application or Functionality? The Type of Information Matters: The Impact of Information about Artificial Intelligence on AI Usage Intention and AI Aversion. https://doi.org/10.31219/osf.io/p3sc2

Saunders PT (2000) Use and abuse of the precautionary principle. ISIS News 6:1474–1547

Schiff KJ, Schiff DS, Adams IT, et al (2023) Institutional factors driving citizen perceptions of AI in government: Evidence from a survey experiment on policing. Public Administration Review puar.13754. https://doi.org/10.1111/puar.13754

Spaccatini F, Richetin J, Riva P, et al (2022) Trust in science and solution aversion: Attitudes toward adaptation measures predict flood risk perception. International Journal of Disaster Risk Reduction 76:103024. https://doi.org/10.1016/j.ijdrr.2022.103024

Stypinska, J. (2023). AI ageism: a critical roadmap for studying age discrimination and exclusion in digitalized societies. *AI & society*, *38*(2), 665-677.

Suleyman M (2023) The coming wave: technology, power, and the twenty-first century's greatest dilemma. Crown

Taylor, S., & Todd, P. (1995). Decomposition and crossover effects in the theory of planned behavior: A study of consumer adoption intentions. *International journal of research in marketing*, *12*(2), 137-155.

Thierer AD (2023) Getting AI Innovation Culture Right. SSRN Journal. https://doi.org/10.2139/ssrn.4404402

West SM, Whittaker M, Crawford K (2019) Discriminating Systems: Gender, Race and Power in AI. AI Now Institute 1–33

Wilson G, Gates JR, Vijaykumar S, Morgan DJ (2023) Understanding older adults' use of social technology and the factors influencing use. Ageing and Society 43:222–245. https://doi.org/10.1017/S0144686X21000490

Yang S, Krause NM, Bao L, et al (2023) In AI We Trust: The Interplay of Media Use, Political Ideology, and Trust in Shaping Emerging AI Attitudes. Journalism & Mass Communication Quarterly 10776990231190868. https://doi.org/10.1177/10776990231190868

Zhang B, Dafoe A (2020) U.S. Public Opinion on the Governance of Artificial Intelligence. In: Proceedings of the AAAI/ACM Conference on AI, Ethics, and Society. ACM, New York NY USA, pp 187–193

# Appendix

*Distribution of responses on AI concerns*

**Concern - Healthcare Provider AI**

| | | | Freq. | Percent | Cum. % |
|---|---|---|---|---|---|
| | 1 | Not at all concerned | 118 | 8.99 | 8.99 |
| | 2 | A little bit concerned | 168 | 12.8 | 21.8 |
| | 3 | Somewhat concerned | 212 | 16.16 | 37.96 |
| | 4 | Moderately concerned | 208 | 15.85 | 53.81 |
| | 5 | Concerned | 178 | 13.57 | 67.38 |
| | 6 | Very concerned | 206 | 15.7 | 83.08 |
| | 7 | Extremely concerned | 222 | 16.92 | 100 |
| Total | | | 1312 | 100 | |

**Concern - Job Threated by AI**

| | | | Freq. | Percent | Cum. % |
|---|---|---|---|---|---|
| | 1 | Not at all concerned | 125 | 9.91 | 9.91 |
| | 2 | A little bit concerned | 107 | 8.49 | 18.4 |
| | 3 | Somewhat concerned | 150 | 11.9 | 30.29 |
| | 4 | Moderately concerned | 184 | 14.59 | 44.89 |
| | 5 | Concerned | 209 | 16.57 | 61.46 |
| | 6 | Very concerned | 219 | 17.37 | 78.83 |
| | 7 | Extremely concerned | 267 | 21.17 | 100 |
| Total | | | 1261 | 100 | |

**Concern - Election Influence**

| | | | Freq. | Percent | Cum. % |
|---|---|---|---|---|---|
| | 1 | Not at all concerned | 51 | 4.08 | 4.08 |
| | 2 | A little bit concerned | 66 | 5.28 | 9.37 |
| | 3 | Somewhat concerned | 102 | 8.17 | 17.53 |
| | 4 | Moderately concerned | 149 | 11.93 | 29.46 |
| | 5 | Concerned | 173 | 13.85 | 43.31 |
| | 6 | Very concerned | 237 | 18.98 | 62.29 |
| | 7 | Extremely concerned | 471 | 37.71 | 100 |
| Total | | | 1249 | 100 | |

**Concern - Manipulated Into Giving Up Info**

|   |   | Freq. | Percent | Cum. % |
|---|---|---|---|---|
| 1 | Not at all concerned | 52 | 4.18 | 4.18 |
| 2 | A little bit concerned | 77 | 6.19 | 10.37 |
| 3 | Somewhat concerned | 93 | 7.48 | 17.85 |
| 4 | Moderately concerned | 163 | 13.1 | 30.95 |
| 5 | Concerned | 184 | 14.79 | 45.74 |
| 6 | Very concerned | 242 | 19.45 | 65.19 |
| 7 | Extremely concerned | 433 | 34.81 | 100 |
| Total | | 1244 | 100 | |

**Concern - Predictive AI Policing**

|   |   | Freq. | Percent | Cum. % |
|---|---|---|---|---|
| 1 | Not at all concerned | 145 | 11.69 | 11.69 |
| 2 | A little bit concerned | 148 | 11.94 | 23.63 |
| 3 | Somewhat concerned | 161 | 12.98 | 36.61 |
| 4 | Moderately concerned | 194 | 15.65 | 52.26 |
| 5 | Concerned | 175 | 14.11 | 66.37 |
| 6 | Very concerned | 174 | 14.03 | 80.4 |
| 7 | Extremely concerned | 243 | 19.6 | 100 |
| Total | | 1240 | 100 | |

**Concern - Children Antisocial**

|   |   | Freq. | Percent | Cum. % |
|---|---|---|---|---|
| 1 | Not at all concerned | 128 | 10.34 | 10.34 |
| 2 | A little bit concerned | 152 | 12.28 | 22.62 |
| 3 | Somewhat concerned | 141 | 11.39 | 34.01 |
| 4 | Moderately concerned | 165 | 13.33 | 47.33 |
| 5 | Concerned | 177 | 14.3 | 61.63 |
| 6 | Very concerned | 186 | 15.02 | 76.66 |
| 7 | Extremely concerned | 289 | 23.34 | 100 |
| Total | | 1238 | 100 | |

**Concern - Robots Become Sentient**

|   |   | Freq. | Percent | Cum. % |
|---|---|---|---|---|
| 1 | Not at all concerned | 150 | 12.14 | 12.14 |
| 2 | A little bit concerned | 141 | 11.41 | 23.54 |
| 3 | Somewhat concerned | 138 | 11.17 | 34.71 |
| 4 | Moderately concerned | 147 | 11.89 | 46.6 |
| 5 | Concerned | 206 | 16.67 | 63.27 |
| 6 | Very concerned | 184 | 14.89 | 78.16 |
| 7 | Extremely concerned | 270 | 21.84 | 100 |
| Total | | 1236 | 100 | |

**Concern - Robots Care for Elderly**

|   |   | Freq. | Percent | Cum. % |
|---|---|---|---|---|
| 1 | Not at all concerned | 210 | 17.03 | 17.03 |
| 2 | A little bit concerned | 177 | 14.36 | 31.39 |
| 3 | Somewhat concerned | 153 | 12.41 | 43.8 |
| 4 | Moderately concerned | 160 | 12.98 | 56.77 |
| 5 | Concerned | 163 | 13.22 | 69.99 |
| 6 | Very concerned | 142 | 11.52 | 81.51 |
| 7 | Extremely concerned | 228 | 18.49 | 100 |
| Total | | 1233 | 100 | |

|  | Mean (Std. Err) | 95% C.I. Lower | 95% C.I. Upper |
| --- | --- | --- | --- |
| Healthcare Provider AI | 4.27 (0.05) | 4.17 | 4.37 |
| Job Threated by AI | 4.56 (0.05) | 4.45 | 4.67 |
| Election Influence | 5.34 (0.05) | 5.24 | 5.44 |
| Manipulated Into Giving Up Info | 5.26 (0.05) | 5.16 | 5.36 |
| Predictive AI Policing | 4.29 (0.06) | 4.18 | 4.40 |
| Children Antisocial | 4.47 (0.06) | 4.36 | 4.59 |
| Robots Become Sentient | 4.42 (0.06) | 4.30 | 4.53 |
| Robots Care for Elderly | 4.00 (0.06) | 3.88 | 4.11 |
| N | 1313 | | |